\documentclass[aip,apl,reprint,groupedaddress]{revtex4-1}
\usepackage{graphicx}
\usepackage{amsmath}
\begin{document}

%\preprint{}

\title{Plasmon-induced demagnetization and magnetic switching in nickel nanoparticle arrays} %Title of paper



\author{Mikko Kataja,$^{1}$ Francisco Freire-Fernandez,$^{1}$ Jorn P. Witteveen,$^{1}$ Tommi Hakala,$^{2}$ P\"aivi T\"orm\"a,$^{2}$ and Sebastiaan van Dijken$^{1}$}
\email[]{sebastiaan.van.dijken@aalto.fi}
%\thanks{}
\affiliation{$^{1}$NanoSpin, Department of Applied Physics, Aalto University School of Science, P.O. Box 15100, FI-00076 Aalto, Finland.}
\affiliation{$^{2}$COMP Centre of Excellence, Department of Applied Physics, Aalto University, FI-00076 Aalto, Finland.}


\date{\today}

\begin{abstract}

We report on the manipulation of magnetization by femtosecond laser pulses in a periodic array of cylindrical nickel nanoparticles. By performing experiments at different wavelength, we show that the excitation of collective surface plasmon resonances triggers demagnetization in zero field or magnetic switching in a small perpendicular field. Both magnetic effects are explained by plasmon-induced heating of the nickel nanoparticles to their Curie temperature. Model calculations confirm the strong correlation between the excitation of surface plasmon modes and laser-induced changes in magnetization.      

\end{abstract}


\maketitle %\maketitle must follow title, authors, abstract and \pacs


Plasmonic nanostructures enable strong local enhancements of the optical field in areas that are substantially smaller than the wavelength of incident light. This capability offers prospects for magnetic recording. In heat-assisted magnetic recording (HAMR), a plasmonic near-field transducer (NFT) reduces the switching field of high-anisotropy materials via local heating.\cite{KRY08,CHA09,STI10} The NFT in this application is a noble metal nanostructure that is placed near the recording medium. Excitation of the NFT at the plasmon resonance frequency efficiently transfers optical energy to a nanoscale region, enabling local switching at reduced magnetic field. In the realm of ultrafast all-optical switching (AOS),\cite{STA07,LAM14} the use of noble metal plasmonic antennas has also been considered. By placing gold antennas on top of a ferrimagnetic TbFeCo film, Liu and co-workers demonstrated the confinement of magnetic switching to sub-100 nm length scales.\cite{LIU15}.

Independent from HAMR and AOS, a new discipline combining plasmonics and magnetism has emerged recently.\cite{TEM10,BEL11} Experiments on pure ferromagnetic metals demonstrate that, despite stronger ohmic damping, they also support surface plasmon resonances.\cite{CHE11,BON11,GRU10} This raises the question if one could nanostructure the magnetic medium itself to trigger magnetic switching via local enhancements of the optical field. To study the effect of plasmon resonances on magnetic switching it is advantageous to consider ferromagnetic nanoparticles of uniform size and shape. In such nanoparticles, plasmon resonances determine the magneto-optical activity via the excitation of two orthogonal electric dipoles.\cite{MAC13} Plasmon resonances in single ferromagnetic nanoparticles are rather broad. Yet, ordering of the particles into a periodic array significantly narrows the spectral response.\cite{KAT15} In this geometry, hybridization between localized surface plasmons and the diffracted orders of the array produces intense surface lattice resonances (SLRs). SLR modes and, thereby, the optical, magneto-optical, and magnetic circular dichroism (MCD) properties of a ferromagnetic nanoparticle array can be tuned by changing the period or symmetry of the lattice\cite{KAT15,KAT16-1,KAT16-2} or the shape of the nanoparticles.\cite{MAC16} These versatile designer tools could thus be exploited to spectrally gauge the significance of the inverse Faraday effect and MCD on all-optical switching in ferromagnetic materials, a topic of intense scientific debate,\cite{LAM14,BER16,COR16,QAI16,ELH16-1,ELH16-2,ELL16,TSE16,GOR16,TAK16,JOH17,CHO17} or to maximize their impact in monochromatic switching experiments. 

\begin{figure}[b]
\includegraphics{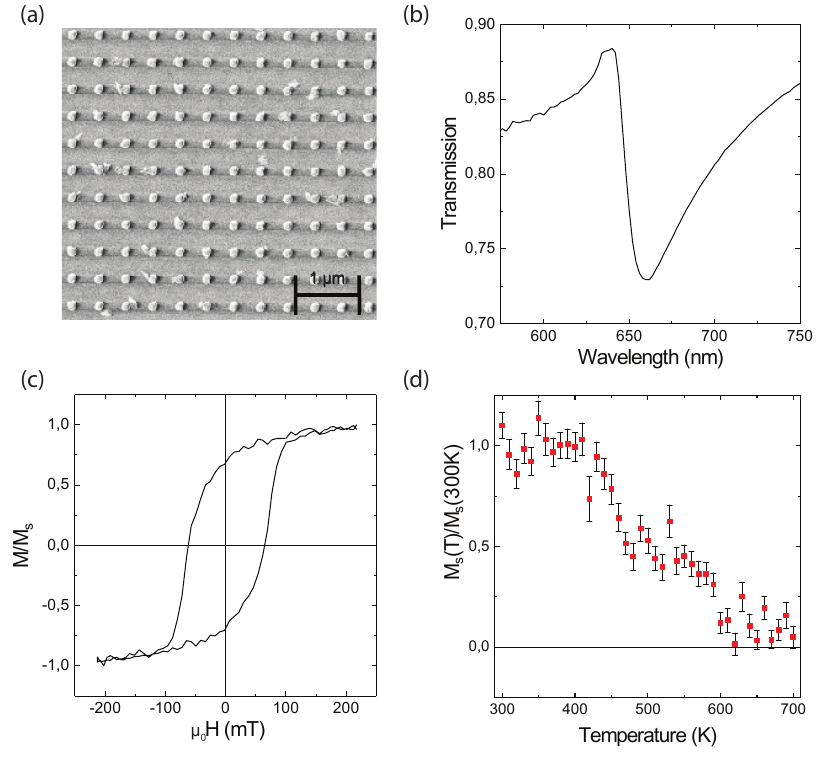}
\caption{\label{Fig1} (a) Scanning electron microscopy image of the nickel nanoparticle array. (b) Optical transmission spectrum. (c) Polar magneto-optical Kerr effect hysteresis loop of the nickel nanoparticle array. (d) VSM measurement of the saturation magnetization as a function of temperature, indicating a Curie temperature of about 600 K.}
\end{figure}

The excitation of surface plasmon resonances in ferromagnetic nanoparticles can cause significant heating. In granular FePt layers where the size, shape, and environment of the ferromagnetic nanoparticles vary, near-field modifications of incident laser radiation have been found to produce an inhomogeneous magnetic switching effect.\cite{GRA17} Here, we study how plasmon resonances affect optical demagnetization and magnetic switching in an ordered array of uniform nickel nanoparticles. We illuminate our sample with femtosecond laser pulses of different wavelength and show how the excitation of a collective SLR mode significantly lowers the pulse fluence that is required to manipulate the magnetization state. Using a phenomenological model, we establish a link between the optical properties and temperature response of the nickel nanoparticle array during femtosecond laser excitation. From this analysis, we conclude that plasmon-induced heating of the nickel nanoparticles to their Curie temperature explains the magnetic effects. 

For this study, we fabricated a periodic array of nickel nanocylinders with a diameter of 100 nm and a thickness of 110 nm on a glass substrate using electron beam lithography (Fig. \ref{Fig1}(a)). The period ($p$) of the square lattice was 420 nm. In all measurements, the nickel nanoparticles were immersed into refractive index matching oil ($n$ = 1.523). Figure \ref{Fig1}(b) shows the transmission spectrum of the array. The spectrum consists of a transmission maximum at the diffracted order of the lattice ($\lambda$ = $np$ = 640 nm). The transmission minimum at $\lambda$ = 660 nm signifies a strong enhancement of optical absorption by the excitation of a SLR mode. 

Figure \ref{Fig1}(c) shows a polar magneto-optical Kerr effect hysteresis loop of the array. The shape of the cylindrical nanoparticles ensures a large perpendicular magnetization in zero magnetic field. The coercive field of the nickel nanoparticles is 62 mT. Using vibrating sample magnetometry (VSM), we measured the dependence of the saturation magnetization on temperature (Fig. \ref{Fig1}(d)). From these data, we estimate a Curie temperature of 600 K, which is only slightly smaller than the transition temperature of bulk nickel.  

\begin{figure}
\includegraphics{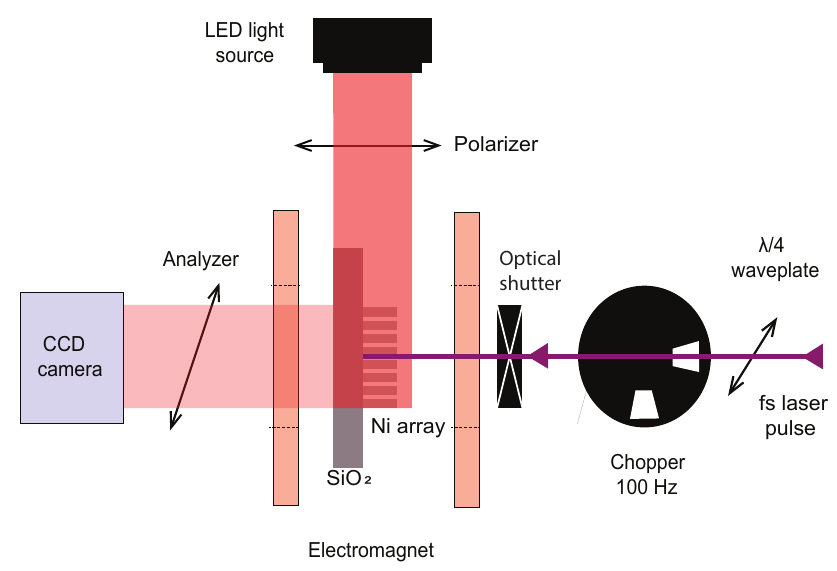}
\caption{\label{Fig2} Schematic of the femtosecond laser setup with dark-field magneto-optical microscope.}
\end{figure}

To study the effect of femtosecond laser pulses on the magnetization of the nickel nanoparticle array we used a Ti:Sapphire laser with an optical parametric amplifier (OPA). We operated the laser at different wavelengths, ranging from 610 nm to 700 nm, in 10 nm intervals. The repetition rate of the laser was 1 kHz. We reduced the frequency to 100 Hz with a chopper and used an optical beam shutter to isolate single pulses with a duration of 100 femtoseconds. We used a dark-field magneto-optical microscopy technique to monitor magnetization changes. In our setup, light from a LED source illuminated the glass sample from the side and a CCD camera detected the scattered light from the nanoparticles. We attained images with good magneto-optical contrast by aligning a polarizer and analyzer at an angle of about 80$^{\circ}$. Figure \ref{Fig2} illustrates the experimental setup.   

\begin{figure*}
\includegraphics{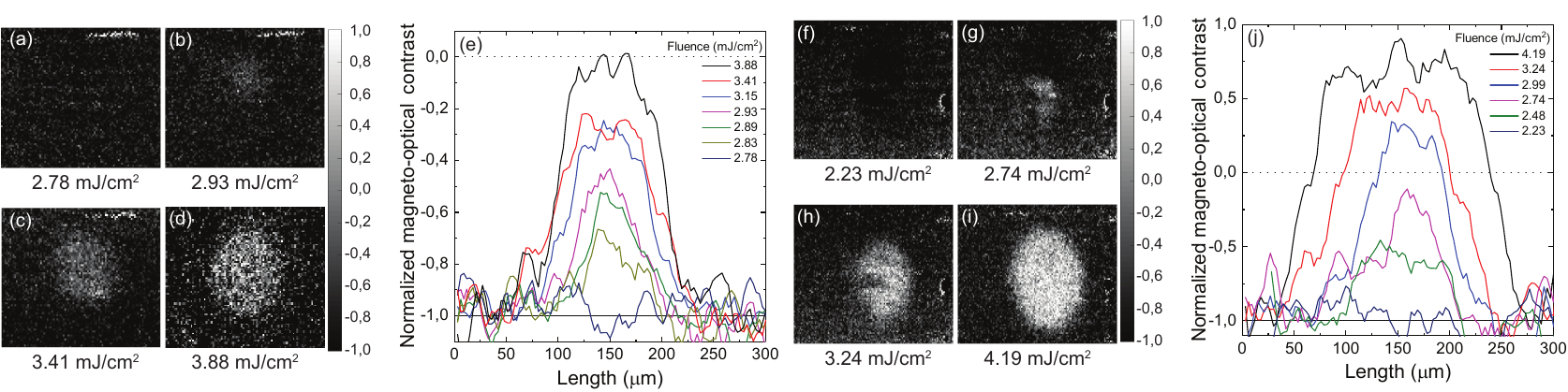}
\caption{\label{Fig3} (a)-(d) Magneto-optical microscopy images of the nickel nanoparticle array after illumination with a single femtosecond laser pulse. The pulse fluence is increased gradually. Femtosecond laser pulses are applied in zero magnetic field and the magnetization is reset after each measurement. (e) Line scans of magneto-optical contrast through the center of the illuminated area. (f)-(i) Magneto-optical microscopy images of the nickel nanoparticle array after illumination with a single femtosecond laser pulse in a perpendicular magnetic field of 5 mT. (j) Line scans of magneto-optical contrast for this configuration. All measurements are performed with a laser wavelength of 660 nm. }
\end{figure*}

Results for femtosecond laser pulses with $\lambda$ = 660 nm are summarized in Fig. \ref{Fig3}. Prior to laser pulse illumination, the magneto-optical contrast was calibrated by full magnetization reversal in a perpendicular magnetic field. After aligning the magnetization along one of the perpendicular directions, we turned off the magnetic field and fired a single laser pulse onto the nickel nanoparticle array. The magneto-optical microscopy images of Fig. \ref{Fig3}(a)-(d) depict the change in magnetization as a function of pulse fluence. We note that the magnetization was reset by an external magnetic field after each laser pulse to avoid accumulative effects. Line scans of magneto-optical contrast through the center of the illuminated area indicate that the magnetization of the array starts to decrease above a pulse fluence of 2.8 mJ/cm$^2$ (Fig. \ref{Fig3}(e)). A fluence of 3.9 mJ/cm$^2$ fully randomizes the magnetization of the nickel nanoparticles in the center of the laser spot. 

In our experiments, we did not find conclusive evidence for all-optical magnetic switching. The use of linearly and circularly polarized laser pulses yielded the same results. To test the possibility of an accumulative MCD effect,\cite{ELH16-1,ELH16-2,ELL16,TSE16,GOR16,TAK16,JOH17} we also illuminated our sample using multiple circularly polarized laser pulses. No evidence for all-optical stochastic switching was found. Magnetic switching could be attained by applying a small magnetic field of 5 mT (only 8$\%$ of the coercive field) during a single femtosecond laser pulse (Fig. \ref{Fig3}(h)-(n)). Under these conditions, a pulse fluence of 4.2 mJ/cm$^2$ reverses the magnetization of most nanoparticles. The absence of a clear dependence on helicity points towards plasmon-induced heating as the source of optical demagnetization and field-assisted magnetic switching. 

To assess the influence of heating, we now wish to establish a link between optical absorption by the nickel nanoparticle array and the required pulse fluence for demagnetization and switching. The extinction of the array, defined as $E=1-T$, is the sum of absorbed and scattered light intensities. Since the extinction cross section ($\sigma_\mathrm{ext}$) of our sample is determined by a SLR mode that absorbs much more than it scatters,\cite{KAT16-1,KAT16-2} we can write $\sigma_\mathrm{ext}=Ep^2\approx\sigma_\mathrm{abs}$. 

\begin{figure}
\includegraphics{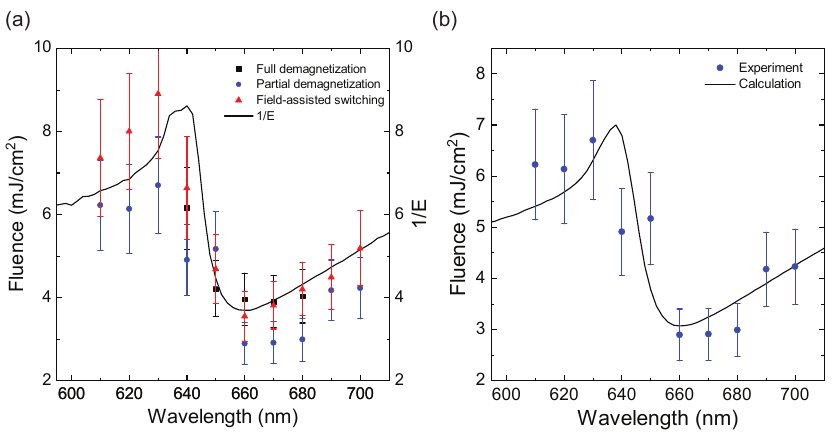}
\caption{\label{Fig4} (a) Experimental threshold fluence for full demagnetization, demagnetization to 0.4$M_\mathrm{s}$ (both in zero magnetic field) and magnetic switching to 0.6$M_\mathrm{s}$ in the opposite direction (magnetic field of 5 mT). The threshold fluences are compared to the inverse extinction (1/$E$) of the nickel nanoparticle array. (b) Experimental threshold fluence (demagnetization to 0.4$M_\mathrm{s}$) and calculation of the fluence that is required to heat the nickel nanoparticles to 600 K.}
\end{figure}

For femtosecond laser pulses, the electronic absorption of optical energy, electron-phonon thermalization and external heat diffusion ensue successively.\cite{BAF13} The total absorbed energy ($Q$) during this process is given by $Q=\sigma_\mathrm{abs}F\approx{Ep^2F}$, where $F$ is the pulse fluence. Since it is reasonable to assume that both demagnetization and magnetic switching require heating of the nickel nanoparticles to their Curie temperature, the threshold pulse fluence ($F_\mathrm{th}$) for both effects should scale as 1/$E$. To test this assertion, we performed similar experiments as those depicted in Fig. \ref{Fig3} using different laser wavelengths. From these measurements, we extracted $F_\mathrm{th}$ for (1) a reduction of the magnetization to 40$\%$ of the initial remanent state, (2) full demagnetization and (3) magnetic switching to 0.6$M_\mathrm{s}$ in a magnetic field of 5 mT. Figure \ref{Fig4}(a) summarizes the results. It also includes the optical transmission data of Fig. \ref{Fig1}(b), now re-plotted as 1/$E$. We find a strong variation of the threshold pulse fluence with laser wavelength. At 660 nm, where the excitation of the SLR mode maximizes the absorption of light, demagnetization and switching by a single femtosecond laser pulse is most efficient. The resemblance between the wavelength dependence of $F_\mathrm{th}$ and 1/$E$ confirms the crucial role of plasmon-induced heating. 

As a final check, we calculate $F_\mathrm{th}$ using a phenomenological model. Just after a single femtosecond laser pulse, the brief and intense temperature increase of a metallic nanoparticle can be estimated as:\cite{BAF11}

\begin{equation}
dT=\frac{Q}{V{\rho}c_\mathrm{p}(T)}.
\end{equation}

Here, $V$, $\rho$ and $c_\mathrm{p}$ are the volume, mass density and specific heat capacity of the nanoparticle. For nickel, the heat capacity varies with temperature.\cite{GAM05} The required energy for heating the nanoparticle from room temperature to a Curie temperature of 600 K can thus be obtained from:

\begin{equation}
Q_\mathrm{CT}=\int_{T=293\mathrm{K}}^{T=600\mathrm{K}}\!\frac{Q}{V{\rho}c_\mathrm{p}(T)}dT 
\end{equation}

After this, the threshold fluence is given by: 

\begin{equation}
F_\mathrm{th}=\frac{Q_\mathrm{CT}}{\sigma_\mathrm{abs}}\approx{}\frac{Q_\mathrm{CT}}{Ep^2}.
\end{equation}

Figure \ref{Fig4}(b) shows the result of this calculation for our cylindrical nickel nanoparticles. Here, we used the experimentally measured extinction as input parameter. The agreement between the measured threshold fluence for demagnetization and the theoretical model indicates that the nanoparticles are indeed heated to their Curie temperature by the femtosecond laser pulse. Thus, the excitation of an intense SLR mode strongly modulates the pulse fluence that is required to manipulate the magnetization of the nickel nanoparticles.  

In summary, we have shown that patterning of a magnetic medium into a periodic nanoparticle array affects thermal demagnetization and heat-assisted magnetic switching via the excitation of collective surface plasmon resonances. Whilst all-optical magnetic switching was not attained in our experiments on nickel, ferromagnetic nanoparticle arrays provide interesting designer tools for further research. In particular, the ability to spectrally separate plasmon-induced heating, the inverse Faraday effect, and MCD via lattice symmetry or nanoparticle shape could offer new insights into the origin of all-optical switching of ferromagnets.    

\begin{acknowledgments}
This work was supported by the Aalto Centre for Quantum Engineering, by the Academy of Finland through its Centres of Excellence Programme (2012-2017) and under project nos. 284621, 303351 and 307419, and by the European Research Council (ERC-2013-AdG-340748-CODE). Lithography was performed at the Micronova Nanofabrication Centre, supported by Aalto University. 
\end{acknowledgments}

%\bibliographystyle{plain}
%\bibliography{references}


\end{document}